# CCD photometry of variable stars in the field of the globular cluster NGC 6397[1]

J. Kaluzny

Warsaw University Observatory, Al. Ujazdowskie 4, 00-478 Warsaw, Poland, e-mail: jka@sirius.astrouw.edu.pl

## ABSTRACT

The $13.6 \times 13.6$ arcmin$^2$ field covering the central part of the globular cluster NGC 6397 was surveyed in a search for short-period variable stars. We obtained light curves for 6 close binaries, 2 SX Phe stars and 1 RR Lyrae variable, 7 of which are new discoveries. Both identified SX Phe variables are likely members of the cluster. One of the identified eclipsing binaries is most probably a field star while for the remaining 5 binaries our data do not preclude cluster membership. A $V$ vs. $B - V$ color-magnitude diagram extending from the tip of the red giant branch to below the main-sequence turnoff is presented.

## 1. Introduction

NGC 6397 is, together with NGC 6101, one of the two nearest globular clusters. In his recent compilation Djorgovski (1993) lists for it $(m - M)_0 = 11.71$, $A_V = 0.56$ and [Fe/H] $= -1.91$. Despite cluster closeness the central part of NGC6397 is relatively difficult to study because of the high surface density of stars in this region. The cluster supposedly went through the core-collapse episode in the past (Djorgovski & King 1986). Alcaino et al. (1987) published photographic photometry for the large part of the cluster as well as CCD photometry for a relatively small area located in an outer part of the cluster. Ground based CCD photometry for the central part of NGC 6397 was published by Auriere et al. (1990) and by Lauzeral et al. (1992). These studies revealed presence of about two dozen blue stragglers in the cluster center. More recently several groups used data obtained with the HST to study stellar population in NGC 6397 (Burgarella et al. 1994, De Marchi & Paresce 1994, Cool et al. 1995, King et al. 1995).

Very few variable stars are known in NGC 6397. The cluster shows a very blue horizontal branch which is void of any RR Lyr pulsators. Hoag (1973) catalogue lists two long period variables belonging to the cluster and one background RR Lyr star. Rubenstein & Bailyn (1993) reported a discovery of 5 SX Phe variables among NGC 6397 blue stragglers. Published light curves of these candidate variables are very noisy and show full amplitudes ranging from a few

---

hundredths of magnitude to 0.15 magnitude. The quoted report had a rather preliminary character and no further details were published so far.

In this contribution we presents results of a mini-survey for short period variables located in the central part of NGC 6397.

## 2. Observations and data reduction

The reported survey was conducted as a supplementary project during a long observing run devoted mainly to a search for eclipsing detached binaries in the globular cluster M4 (Kaluzny 1996). The central part of NGC 6397 was monitored with the CTIO 0.9-m telescope and Tektronix 2048 No. 3 CCD. The field of view of the camera was $13.6 \times 13.6$ arcmin$^2$ with scale of 0.396 arcsec/pixel. Observations were performed using the Johnson $B$ and $V$ filters. The exposure time ranged from 150 to 300 sec for the $B$-band, and from 90 to 200 sec for the $V$-band. To search for variable stars we used 83 frames in $B$ and 142 frames in $V$. All the $B$-band exposures and most $V$-band exposures were collected on the nights of July 8/9 and July 9/10, 1995. A few additional exposures in the $V$-band were collected on three subsequent nights. A condensed log of observations is given in Table 1. A more detailed log was submitted to the editors of A&A (see Appendix A). Frame numbers quoted occasionally below refer to that log.

The preliminary processing of the raw data was made with IRAF [2]. The flat-field frames were prepared by combining sets of 10-15 frames obtained by observing an illuminated screen in the dome. The reduction procedures reduced total instrumental systematics to below 1% for the central $1500 \times 1500$ pixels$^2$ area of the images. Some systematic residual pattern at the 1%-2% level was left near borders of the images.

Stellar profile photometry was extracted using DoPHOT (Schechter et al. 1993). We used DoPHOT in the fixed-position mode. The stellar positions were provided based on a list obtained by reduction of a "template" image. Two consecutive images of a good quality (frames #1703 and #1704; both with FWHM = 1.25 arcsec and $t_{\exp}$ = 100 sec) were combined to produce template for the $V$ filter observations. An individual image (frame #1705, FWHM = 1.52 arcsec, $t_{\exp}$ = 180 sec) served as a template for observations in the $B$ filter.

The images collected with the CTIO 0.9-m telescope show a significant positional dependence of the point spread function. To cope with this effect we applied a procedure similar to that described in details in Kaluzny et al. (1996). In short, each analyzed frame was divided into a $5 \times 5$ grid of overlapping sub-frames. An instrumental photometry derived for a give sub-frame of the frame was transformed to the common instrumental system of the "template" image. The data bases for the $V$ and $B$ filers contained light curves for 12259 and 17654 stars, respectively.

---

[2]IRAF is distributed by the National Optical Astronomy Observatories, which are operated by the Associations of Universities for Research in Astronomy, Inc., under cooperative agreement with the National Science Foundation

## 2.1. The color-magnitude diagrams

Photometry derived from "template" frames was used to construct the color-magnitude diagram (CMD) for the monitored field. The instrumental photometry was extracted using DAOPHOT/ALLSTAR (Stetson 1987). Following Walker (1994) we selected a Moffat-function point spread function, quadratically varying with X and Y coordinates. The instrumental photometry was transformed to the standard BV system using relations:

$$v = \text{const} + V - 0.0102 \times (B - V) \qquad (1)$$
$$b - v = \text{const} + 1.1376 \times (B - V) \qquad (2)$$

where lower case letters correspond to the instrumental magnitudes. The color terms of the transformation were determined based on observations of standard stars from the Landolt (1993) fields. The zero points were calculated using photometry of 9 local standards from NGC 6397 field (Alcaino & Liller 1986). The derived CMD of the cluster is shown in. Fig. 1. Poor measurements were flagged in the photometry files generated by DAOPHOT/ALLSTAR. A given measurement was considered to be poor when the formal error of photometry was 2.5 times or more larger than the average error of photometry for other stars of comparable magnitudes. Stars for which either $V$ or $B$ photometry was flagged as poor, were not plotted in Fig. 1. Stars with $V < 12.0$ were in fact badly overexposed on the template image for the $V$ filter. Their $V$ photometry was extracted from a single, shortly exposed image (frame #1911, FWHM=1.85, $t_{\exp} = 20$ sec).

Another CMD of the monitored field was constructed based on average magnitudes of stars included in the data bases for both filters. To calculate average magnitudes we selected only relatively good frames ($FWHM < 1.68$ arcsec, clear sky during observation). A total of 49 and 97 frames were selected for $B$ and $V$ bands, respectively. The average magnitudes were then calculated for stars with at least 15 and 35 measurements in the $B$ and $V$ bands, respectively. The 3 most strongly deviating measurements were rejected while calculating average magnitudes. Moreover, we used only measurements to which DoPHOT assigned type equal 1 (stellar objects of relatively high S/N). The resultant CMD is presented in Fig. 2. This photometry covers a smaller range of magnitudes than that presented in Fig. 1. However, the CMD based on average magnitudes has a better internal accuracy and shows the better defined main-sequence of the cluster.

The tables with photometry presented in Figs. 1&2 are available in the electronic form from the editors of A&A (see Appendix A). We emphasize that the presented CMDs were obtained as a by-product of a survey for variable stars. These observations were not aimed of getting a deep and accurate photometry suitable for study of the cluster properties. All images of the cluster were obtained under a rather unfavorable conditions (see Table 1). The zero points of the presented BV photometry are based entirely on the local standards (Alcaino & Liller 1986). These local standards were set up using a photoelectric photometr.

## 3. Results for variable stars

The search for potential short-period variables was conducted by analyzing light curves included in data bases for the $B$ and $V$ filters. To select potential variables we employed two methods which are described in details by Kaluzny et al. (1996). The first method is based on selecting stars exhibiting "noisy" light curves. An $\chi^2$ test is applied to the light curves to identify such stars. In the second method the light curves are scanned with a filter designed to detect an eclipse-like events. To illustrate the overall quality of photometry we present in Fig. 3 a plot of $rms$ deviation versus the average $V$ magnitude for the light curves containing at least 60 data points.
Seven certain variables were identified this way. The data collected on the night of 8/9 July 95 are of superior quality as compared with the remaining nights. Therefore, these data were subject to an additional analysis. As a result one more variable (star V11 in Table 2) was identified. This variable shows a light curve with a very low amplitude and it is located very close to the projected cluster center.

The rectangular coordinates and basic photometric properties of the identified variables are listed in Table 2. The listed rectangular coordinates correspond to the positions of variables on the "template" V-band image which was submitted to editors of A&A (see Appendix A).

The periods of pulsating variables were determined using the *aov* statistic (Schwarzenberg-Czerny 1989, 1991). Several moments of individual minima were obtained for eclipsing binaries V4-9. These moments are listed in Table 3. Because of the very short time-base of our observations all periods listed in Table 2 are of low accuracy. In particular, the period of V6 is just a rough estimate derived from an incomplete light curve. However, we exclude a possibility of any gross errors caused by aliasing related problems.

Variable V3 is listed in the Hoag (1973) catalogue as a background object with the period P=0.330667 d. We revised the period of V3 to P=0.494 d. Variable V10 is most probably identical to the variable blue straggler BSS #533 reported by Rubenstein and Bailyn (1993). This conclusion is based on a close similarity of periods, light curve amplitudes and magnitudes reported for V10 and BSS #533. Variables V4-9 and V11 are, to the best of our knowledge, new discoveries.

The phased light curves of variables V3-4, V7-9 and V11 are shown in Fig. 4. A time-domain light curves of V5, V6 and V10 are presented in Fig. 5. Only the $V$-band light curves are displayed in this paper. The tables containing $B$ and $V$ light curves for all the variables discussed here are available from A&A (see Appendix A).

## 4. Properties and cluster membership of the variables

Figure 6 shows a CMD of NGC 6397 with marked positions of 8 out of 9 variables detected in our survey. Positions of pulsating variables correspond to their average magnitudes and colors. For RR Lyr variable V3 we obtained intensity averaged magnitudes $<V> = 16.07$ and $<B> = 16.54$. This star occupies a position about 3 magnitudes below the horizontal branch of NGC 6397 and it is clearly a background object. For SX Phe stars we approximated average magnitudes with the arithmetic mean of $V_{max}$ and $V_{min}$. We obtained $<V> = 15.80$ and $<V> = 15.37$ for V10 and V11, respectively. Both SX Phe stars detected in our survey are located among candidate blue stragglers on the cluster CMD. Let us assume for a moment that V10 and V11 are members of the cluster. Adopting for NGC 6397 $(m-M)_V = 12.27$ (Djorgovski 1993) we obtain $M_V^{obs} = 3.53$ and $M_V^{obs} = 3.10$ for the observed absolute magnitudes of V10 and V11, respectively. On the other hand, we may use the period-luminosity relation to estimate absolute magnitudes of SX Phe stars. According to criteria introduced by McNamara (1995) V10 and V11 are the first-overtone pulsators. We adopted the relation $M_V = -3.29 \times \log P - 1.53$ (McNamara 1995) and obtained $M_V^{cal} = 3.46$ and $M_V^{cal} = 3.12$ for V10 and V11, respectively. A very good agreement between the observed and calculated values of $M_V$ indicates that the both SX Phe stars are likely members of the cluster.

We discovered one probable and four certain contact binaries in the cluster field. Systems V4 and V9 are located about 0.3 mag to the red of the cluster main sequence. Binary V7 is located close to the cluster turnoff among candidates for faint blue stragglers while V8 occupies a position only slightly above the main sequence. We have applied the absolute brightness calibration established by Rucinski (1995) to calculate $M_V^{cal}$ for newly discovered contact binaries. Assuming that all systems are members of the cluster we adopted for them [Fe/H] = $-1.91$ and $E(B-V) = 0.18$. Values of $M_V^{obs}$ were derived by adopting again $(m-M)_V = 12.27$ for all systems. The derived values of $M_V^{cal}$ and $M_V^{obs}$ are given in Table 4. The presented results, if taken strictly, indicate that V4 and V9 are background objects while V7 and V8 are located in the cluster foreground. On the other hand, the differences between $M_V^{obs}$ and $M_V^{cal}$ are relatively small. The Rucinski's (1995) calibration is rather preliminary in respect of metal poor systems and the calibration stars exhibit a scatter with $\sigma \approx 0.3$ mag around the adopted relation. Therefore, the available data do not allow to reject a hypothesis about cluster membership of any of newly discovered contact systems.

We excluded from the above discussion the variable V6 for which only a rough estimate of a period is accessible for a moment. This star is located on the CMD far away to the red from the cluster main-sequence. Most probably V6 is not a member of NGC 6397.

We have failed to determine the $B-V$ color for variable V5. Therefore its position on the cluster CMD is unknown. This star was positioned very closely to the edge of the monitored field. Its photometry was difficult also because of presence of 2 close and bright companions. A visual inspection of several frames confirmed beyond any doubts a variability of V5. The light curve of this star shows two distinct minima. Because of the noisy photometry it is difficult to prove or disprove a hypothesis that V5 is a contact binary. We may only note that almost all known eclipsing main-sequence binaries with periods shorter than about 0.4 day show EW-type light curves. The light curve of V5 is very atypical in this respect. Further observations are needed to clarify status of V5.

## 5. Blue stragglers

As we noted in the Introduction, NGC 6397 is known to harbor a large number of blue stragglers. We examined in detail light curves of 54 candidates for blue stragglers selected based on CMD presented in Fig. 2. Specifically, we selected stars with:
a) $14.0 < V < V15.6$ and $(B - V) < 0.6$ or
b) $15.6 < V < 16.0$ and located to the blue of the line connecting points $(B - V, V) = (0.52, 16.0)$ and $(B - V, V) = (0.60, 15.60)$. This sample included 2 earlier detected SX Phe stars. For 47 stars we could rule out any periodic variability with periods $P < 3$ hours and the full amplitude exceeding about 0.04 magnitude. Hence, the relative frequency of SX Phe stars among NGC 6397 blue stragglers is higher or equal to 2/49. Note that some candidates for blue stragglers can in fact be field stars, which would increase the estimated relative frequency of occurrence of SX Phe stars. Moreover, an example of V11 shows that some SX Phe stars can exhibit light curves with very small amplitudes.

## 6. Discussion and Summary

The main result of this paper is the demonstration that it is feasible to survey NGC 6397 for eclipsing binaries using the 1-m class telescope under mediocre weather conditions. Detection of detached eclipsing binaries among upper main-sequence stars in NGC 6397 would be of great interests. By combining radial velocity curves with photometry one would be able to determine absolute parameters for components of such binaries. This in turn would provide information about age of the cluster.

The periods of two SX Phe stars identified in NGC 6397 confirm the correlation noted by McNamara (1995). He noted that the shortest period variables are found preferably in the most metal-poor clusters. In fact star V10 has the shortest period among all known SX Phe stars.

We identified six close binaries in the surveyed field. Four of them are probable cluster members and one is most probably a field object. The membership status and type of the sixth variable cannot be determined based on our photometry.

As a by product of our mini-survey we obtained two sets of BV photometry for stars from the central part of the cluster. The more extended set contains data for 13028 stars. The second set includes 9852 stars but the reduced number of objects is compensated by the relatively higher quality of the photometry. Although the derived photometry is not very deep as for temporary standards it does cover much larger area than any other CCD based study published so far for NGC 6397.

This project was supported by the Polish KBN grant 2P03D00808 to JK and by NSF grant AST 93-13620 to Bohdan Paczynski. Thanks are due to Kazik Stępień for comments on the draft version of this paper.

## 7. Appendix A

Tables containing light curves of all variables discussed in this paper as well as tables with BV photometry presented in Figs. 1& 2 are published by A&A at the Centre de Données de Strasburg, where they are available in the electronic form: See the Editorial in A&A 1993, Vol. 280, page E1. We have also submitted the V-filter "template" image of the monitored field. This image can be used for identification of all variables discussed in this paper as well as for identification of all stars for which we provide BV photometry. A table containing detailed log of all observations used in this paper is also available from A&A.

# REFERENCES

Alcaino, G., & Liller, W. 1986, AJ 91, 89

Alcaino, G., et al. 1987, AJ 94, 917

Auriere, M., Ortolani, S., Lauzeral, C. 1990, Nat 344, 638

Burgarella, D. et al. 1994, A&A 287, 769

Cool, A.M, Grindley, J.E., Cohn, H.N., Lugger, P.M., Slavin, S.D. 1995, ApJ 439, 695

De Marchi, G., & Paresce, F. 1994, A&A 281, L13

Djorgovski, S. 1993, APS Conf Ser., Vol. 50, eds. S.G. Djorgovski & G. Meylan, p. 373

Djorgovski, S., & King, I.R. 1986, ApJ 305, L61

Hoag, H.S. 1973, Publ. DDO 6, No. 3, p. 1

Kaluzny, J. 1996, in preparation

Kaluzny, J., & Krzeminski, W. 1993, MNRAS, 264, 785

Kaluzny, J., Kubiak, M., Szymański, Udalski, A., Krzeminski, W., & Mateo, M. 1996, submitted to A&A

King, I.R., Sosin, C., & Cool, A.M. 1995, ApJ 452, L33

Landolt, A.U. 1992, AJ, 104, 340

Lauzeral, C., Ortolani, S., Auriere, M., & Melnick, J. 1992, A&A 262, 63

Mateo, M. 1995, in "Binaries in Clusters", ASP Conf. Ser., eds. G. Milone and J.-C. Mermilliod, in print

McNamara, D.H. 1995, AJ 109, 1751

Nemec, J.M., Linnell Nemec A.F., Lutz, T.E. 1994, AJ 108, 222

Rubenstein, E.P., & Bailyn, C.D. 1993, ASP Conf. Ser., Vol. 50, ed. R.A. Saffer, p.357

Rucinski, S.M. 1995, PASP 107, 648

Schechter, P., Mateo, M., & Saha, A. 1993, PASP 105, 1342

Schwarzenberg-Czerny, A. 1989, MNRAS 241, 153

Schwarzenberg-Czerny, A. 1991, MNRAS 253, 198

Stetson, P.B. 1987, PASP 99, 191

Walker, A.R. 1994, AJ 108, 555

---

Table 1: Log of observations of NGC 6397. The third column gives a number of frames collected in both filters.

| Date(UT) | duration (hours) | N | Seeing arcsec | Remarks |
|---|---|---|---|---|
| 8/9 July, 95 | 7.5 | 75 | 1.2-1.8 | Full Moon, clear |
| 9/10 July, 95 | 9.0 | 56 | 1.3-1.8 | Cirrus |
| 12/13 July, 95 | 1.3 | 7 | 1.6-1.9 | Cirrus, $V$ only |
| 13/14 July, 95 | - | 2 | 2.9 | Cirrus, $V$ only |
| 14/15 July, 95 | - | 2 | 2.7 | Cirrus, $V$ only |

Table 2: Rectangular coordinates and basic photometric data for variables V3-12 from the field of NGC 6397. Uncertain quantities are marked with a colon. The periods are given in days and $\delta V$ is a full amplitude of variability in the V-band.

| Name | X | Y | $V_{max}$ | $(B-V)_{max}$ | $\delta V$ | P | Type |
|---|---|---|---|---|---|---|---|
| V3 | 1538.64 | 1131.24 | 15.32 | 0.26 | 1.20 | 0.49405 | RRab |
| V4 | 390.84 | 1354.07 | 16.26 | 0.89 | 0.74 | 0.4218 | EW |
| V5 | 465.62 | 12.01 | 18.70 | | > 0.8 | 0.27: | EA/EB: |
| V6 | 732.61 | 1517.37 | 17.16 | 1.17 | 0.13: | 0.5: | EW: |
| V7 | 942.84 | 1058.19 | 16.80 | 0.61 | 0.38 | 0.27165 | EW |
| V8 | 1046.84 | 788.94 | 16.19 | 0.50 | 0.37 | 0.27100 | EW |
| V9 | 1869.78 | 339.85 | 16.25 | 0.91 | 0.15 | 0.58017 | EW |
| V10 | 1083.94 | 1061.60 | 15.78 | 0.35: | 0.11: | 0.0301 | SX Phe |
| V11 | 938.18 | 1071.69 | 15.32 | 0.34: | 0.04: | 0.0382 | SX Phe |

Table 3: Moments of minimum light determined from the V-band light curves of variables V4-9.

| ID | $T_{min}$ (HJD) 2449000+ | $\sigma$ | E |
|---|---|---|---|
| V4 | 8.5982 | 0.0003 | 0 |
| V4 | 9.6543 | 0.0007 | 2.5 |
| V5 | 8.5862 | 0.0005 | 0.5 |
| V5 | 8.7179 | 0.0006 | 1.0 |
| V6 | 8.646 | 0.001 | 0.5 |
| V7 | 8.5044 | 0.0006 | 0. |
| V7 | 8.6396 | 0.0007 | 0.5 |
| V8 | 8.5986 | 0.0002 | 0.5 |
| V8 | 9.6823 | 0.0005 | 4.5 |
| V9 | 8.6893 | 0.0007 | 0.5 |
| V9 | 9.5613 | 0.0012 | 2.0 |

Table 4: Calculated and observed absolute magnitudes of contact binaries from the field of NGC 6397. Both quantities were derived under the assumed cluster membership of variables.

| ID | $M_V^{cal}$ | $M_V^{obs}$ |
|---|---|---|
| V4 | 4.77 | 3.99 |
| V7 | 4.03 | 4.53 |
| V8 | 3.57 | 3.92 |
| V9 | 4.53 | 3.98 |

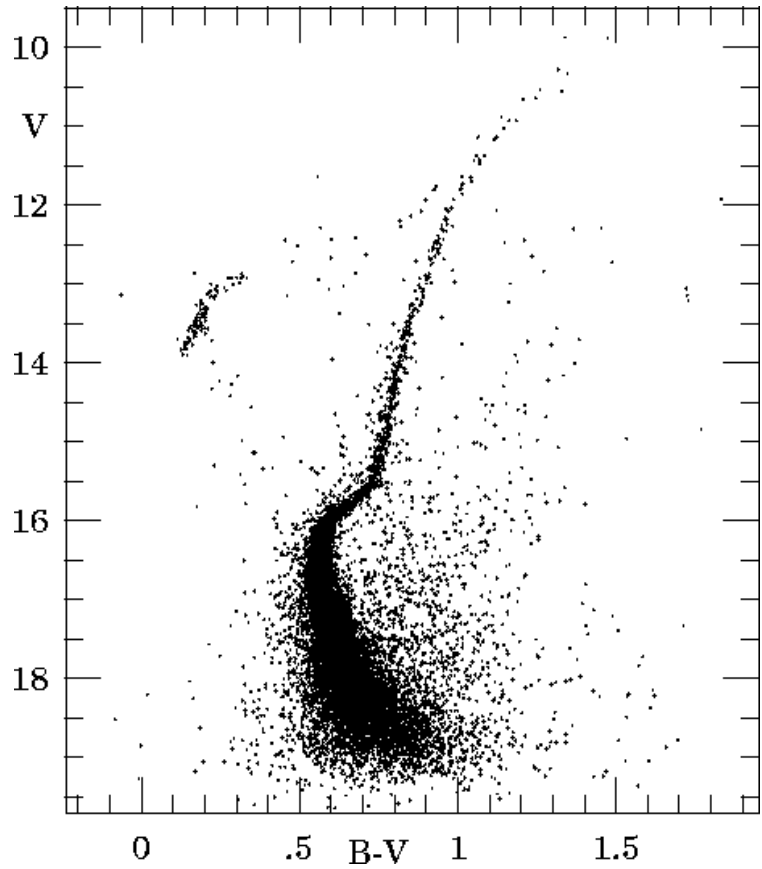

Fig. 1.— The color-magnitude diagram for the monitored field. The presented photometry was extracted from a pair of $BV$ frames with the help of DAOPHOT. The $V$-band photometry for the brightest stars was extracted from an additional shortly exposed image. Stars with unreliable photometry and $V > 14.0$ were not plotted.

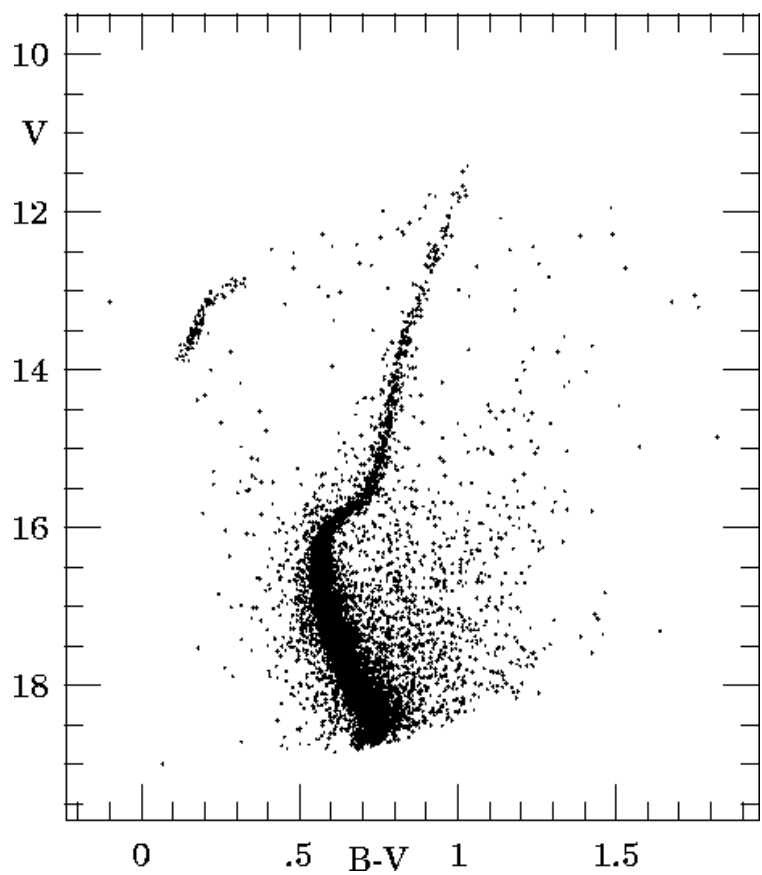

Fig. 2.— The color-magnitude diagram based on the averaged photometry derived with the help of DoPHOT.

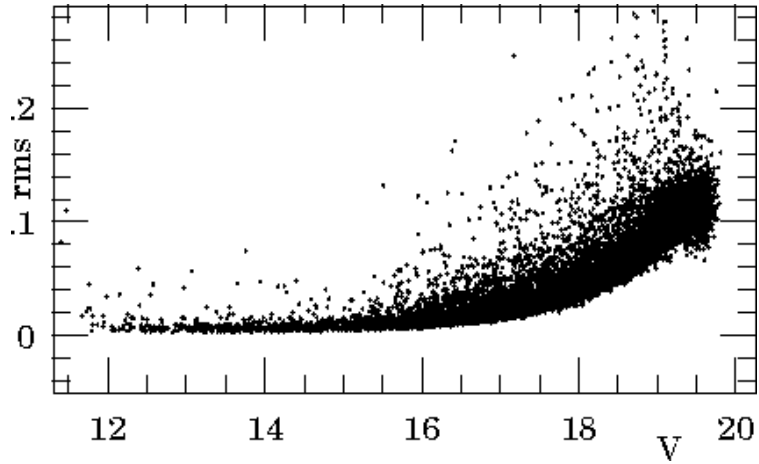

Fig. 3.— The single-measurement errors of our photometry versus the average $V$ magnitudes for 16177 stars whose light curves were examined for variability.

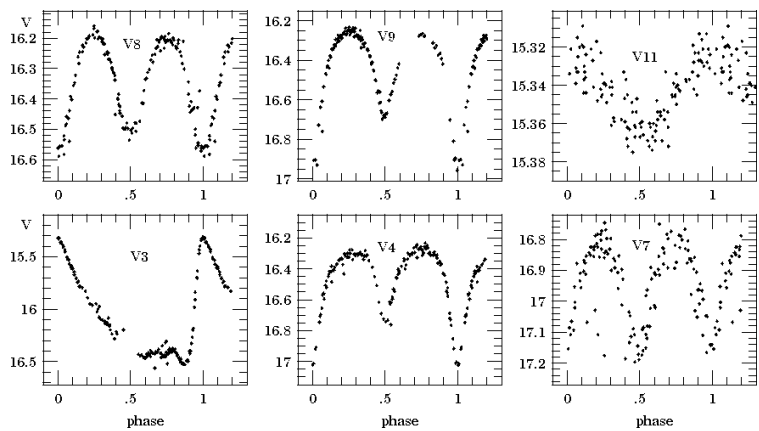

Fig. 4.— Phased $V$-band light curves of variables V3-4, V7-9 and V11.

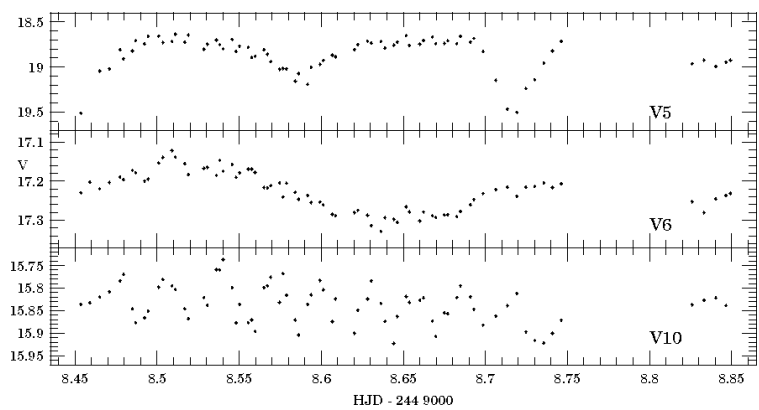

Fig. 5.— Light curves of variables V5, V6 and V10 obtained on the night of July 8/9, 1995

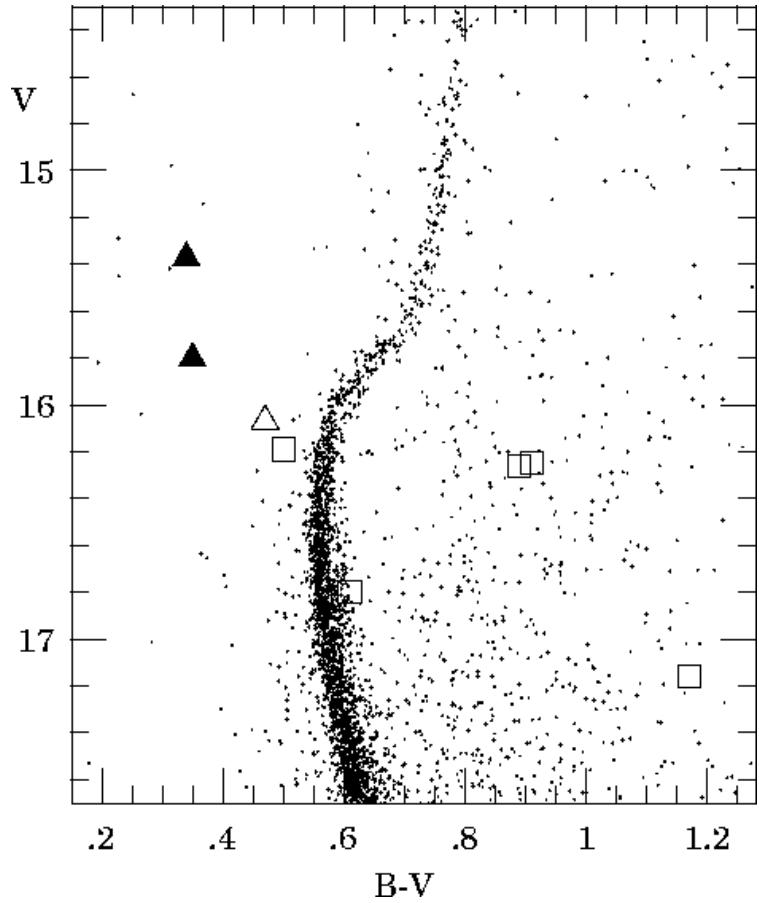

Fig. 6.— The color-magnitude diagram of NGC 6397 with the positions of the variables. Squares correspond to certain or probable (eg. V6) eclipsing binaries, filled triangles to SX Phe stars and an open triangle to RR Lyr star V3. For the sake of clarity, stars from the most crowded part of the observed field are not marked.